# The Futility of Utility

*How market dynamics marginalize Adam Smith*

Joseph L. McCauley
Physics Department
University of Houston
Houston, Texas 77204

## Abstract

Econometrics is based on the nonempiric notion of utility. Prices, dynamics, and market equilibria are supposed to be derived from utility. Utility is usually treated by economists as a price potential, other times utility rates are treated as as Lagrangians. Assumptions of integrability of actions and dynamics are implicitly and uncritically made. In particular, economists assume that price is the gradient of utility in *equilibrium*, but I show instead that price as the gradient of utility is an *integrability* condition for the Hamiltonian dynamics of an optimization problem in econometric control theory. One consequence is that, in a nonintegrable dynamical system, price cannot be expresssed as a function of demand or supply variables. Another consequence is that utility maximization does not describe equilibrium. I point out that maximization of the Gibbs entropy of the observed price distribution of an asset *would* describe equilibrium, *if* equilibrium could be achieved, but equilibrium does not describe real markets. To emphasize the inconsistency of the economists' notion of 'equilibrium', I discuss both deterministic and stochastic descriptions of the dynamics of excess demand and observe that Adam Smith's stabilizing hand is not to be found either in deterministic or stochastic dynamical models of markets, nor in the observed motions of asset prices. Evidence for stability of prices of assets in free markets simply has not been found.

## 0. Overview

Section 1 defines utility, marginal utility and 'General Equilibrium Theory' (GET), which postulates price as the gradient of utility as an equilibrium condition. 'Equilibrium' is believed by economists to be defined by utility maximization.

Part 2 introduces the dynamics of GET, the Arrow-Block-Hurwicz model (ABH), and the search for stable equilibria in econometrics. Motion toward a unique stable equilibrium point, if it occurs, represents the universal action of Adam Smith's hand. Scarf showed that it is easy to construct models where stable equilibria do not occur. Saari pointed out that chaos is not excluded. *I add the observation that deterministic excess demand dynamics also does not exclude complexity.* In any case, Adam Smith's invisible hand is not there to match supply to demand in unregulated free markets. Parts 1 and 2 form the background for part 3.

*Part 3 contains the first main result of the paper.* There, I discuss the Hamiltonian systems that arise from the use of optimization and control theory in econometrics. I point out that the condition that price is a gradient of utility is an integrability condition for the dynamics, not an equilibrium condition, that utility is generally a functional, not a function, and that stable equilibria cannot occur at all in this context.

Part 4 follows the pioneer Osborne in discussing markets and trading via a toy model of limit orders, and follows Black in observing that noise is essential for liquidity. Here, I emphasize the behavior of excess demand in real, liquid markets in favor of the econometric notions and models of parts 1-3, like price as a function of demand and ABH dynamics, that are empirically inapplicable to real markets.

Section 5 introduces Osborne's lognormal model of noisy stock prices. The CAPM (Capital Asset Pricing Model) is derived from white noise. This sets the stage for the simplest possible derivation of the Black-Scholes equation in part 6, and also for criticizing the economists' notion of 'equilibrium' in part 8.

In section 6, following Black, the Black-Scholes option pricing model is derived from the CAPM. It is observed that the notion of utility (along with Adam Smith's hand) is not used in modeling stock prices and apparently can be forgotten. Parts 4-6 form the background for part 8.

Section 7 explains why the study of prices as deterministic flows in the ABH model and control theory, even in the chaotic or complex case, cannot model the observed random noise in stock prices at short times.

*Part 8 contains the second main result of the paper*. There, I note that there is no idea of stability or equilibrium in the stochastic dynamics used to model asset prices, and that neither the lognormal nor the observed (exponential) empirical distribution defines an 'equilibrium' price that characterizes the market. Adam Smith's hand is observed to have disappeared altogether, because there are no restoring forces: excess demand is described as pure random noise. The economists mislabel stochastic dynamics as 'equilibrium dynamics' because they maximize expected utility (which does *not* guarantee equilibrium) instead of maximizing entropy which would, if equilibrium were possible. *I define entropy as the Gibbs entropy of the observed price distribution. The observation that markets are nearly never in equilibrium runs counter to Black's stated belief that equilibrium always prevails in the long run*. I also speculate on what Black may have meant by 'equilibrium'.

Part 9 informs the reader of Osborne's observation that the supply-demand 'graphs' in Samuelson's text are merely cartoons that have not been, and cannot be, derived from empirical data.

Part 10 summarizes and concludes with the expectation that econometrics in its present form cannot survive.

**1. What is econometrics?**

The prevailing notion of western econometrics, expressed in standard texts, begins with Adam Smith's idea of an 'invisible hand' that is supposed somehow to provide stability and fulfill wants optimally by matching supply to demand in free, unregulated markets. Since smith did not and could not describe how this was supposed to work, it has been the main job of econometrics to try to mathematize the idea. The idea of static equilibrium, borrowed from physics, plays a main role in econometric theorizing. An equilibrium market is defined [1,2] as one where the match of supply to demand determines the price of a commodity. In such a market trades are imagined to take place only at the equilibrium price. In order to mathematize, the notion of 'utility' is invented. The following mathematics, which I will call econo-logic, is thereby postulated as the underpinning of General Equilibrium Theory (GET). The main purpose of part 4 of this paper is to discuss a simplified example of a real, liquid market where this idea fails completely, the stock market. First, in parts 1-3, I will review the economists' reasoning and add some new, more precise criticism than has been provided by others, including Black (see part 10 below).

Given n commodities in quantities $x_1,....x_n$, a function $U(x_1,...x_n)$, called 'satisfaction' (economists call it 'utility') is postulated. Utility is supposed to describe consumer preferences, but only for a 'rational' consumer. A rational consumer is one who maximizes utility subject to a budget constraint, completing the circle. The function U is assumed to be concave, building in the assumption of decreasing returns: less satisfaction from eating five ice-cream cones than from eating one. Because utility functions cannot be derived without more information, one must consider the entire class of concave utility functions. *The attempted classification is topologically well-defined only locally, not globally, because it is not invariant under arbitrary nonlinear coordinate transformations.*

Equilibrium is described as the result of maximizing U subject to a budget constraint [1]

$$M = \tilde{p}x = \text{constant} \quad (1)$$

where x is the state vector in commodity space with entries $x_i$ and p is a covector ("co-state vector") whose $i^{th}$ entry is the price $p_i$ of the $i^{th}$ commodity. I have used matrix notation

$$\tilde{p}x = (p_1 \ldots p_n)\begin{pmatrix} x_1 \\ x_n \end{pmatrix} = p_1 x_1 + \ldots + p_n x_n \tag{1b}$$

but sometimes below will use Gibbs notation

$$\nabla_p \bullet \varepsilon(p) = \sum_{i=1}^{n} \frac{\partial \varepsilon_i}{\partial p_i} \tag{1c}$$

to denote inner products of covectors with vectors. Maximizing the utility subject to the budget constraint (1) yields

$$p_i = \lambda \frac{\partial U}{\partial x_i}, \tag{2}$$

where $\lambda$ is a Lagrange multiplier. The utility slopes along the various Cartesian axes are called marginal utilities. According to economists, equilibrium prevails whenever

$$\frac{\text{marginal satisfaction of eating beans}}{\text{price of N beans}} = \frac{\text{marginal satisfaction of owning a BMW}}{\text{price of a BMW}} = \ldots = \lambda \tag{3}$$

An old assumption is to try to identify $\lambda$ as the marginal utility of money.

It's not clear that (3) correctly describes eating beans: eating five beans is more satisfying than eating one bean, if you like beans, although eating ten thousand beans at one sitting, compared with eating ten beans, probably is not. This math-mythology also does not describe stock market bubbles where higher prices invite still higher prices (increasing returns) until the crash. Human satisfaction is surely more complex than is assumed within the confines of econo-logic (neo-classical economics), but let us not become mired in less fruitful objections at this point. Fundamental, fatal flaws in the econometric definition of 'equilibrium' will be analyzed in parts 2, 3, and 8 below. Toward that end I begin historically, motivated by Mirowski's (stimulating but incomplete) criticism [3] of the asumption that utility exists as a function of demand or supply variables.

There is a long-standing internal inconsistency in GET: given a price covector p(x) (price as a function of supply or demand x) the integral

$$A_r = \int_C \tilde{p} dx \quad (4)$$

typically depends on the integration path C. Utility is defined by integrating a nonintegrable differential form [3,4], so that utility is more like 'heat', or a Lagrangian, than potential energy or free energy. Utility is a path-dependent functional, not a function, unless an integrability requirement is satisfied by the differential form $\tilde{p}dx$. The integrability condition is generally not satisfied, but this has not stopped econometricians from assuming utility *functions* in their model-making. Samuelson [4] tried to argue the integrability problem away by suggesting that if consumers would only behave in a certain way then utility would be path-independent. This is a good example of econo-logic, but it would be like trying to change the empirical data to fit the theory rather than the reverse, if empirical data had been used. The general problem of nonintegrability has been faced at an elementary level but was never solved within econometrics. Typically, integrability has been assumed at all levels in econometrics whenever it was convenient to do so, including in dynamical generalizations of GET (see especially part 3 below).

Adam Smith's invisible hand does not have a chance to act in the hypothetical trading that occurs *at equilibrium* in GET. The hand did not bring prices together because nothing was done to reach equilibrium (no dynamics were involved). To remedy this dynamics-gap Walras suggested that we imagine that an auction has occured in order to reach equilibrium. Adam Smith's invisible hand is then a synonym for the bidding and price adjustments during the bidding. Imagine many traders at a market, each with his own initial commodity vector $x_o$ held with initial price covector $p_o$. An auctioneer, called the Walras auctioneer, calls out a price vector p. Each trader computes his excess demand

$$\varepsilon_1(p) = (\tilde{p} - \tilde{p}_o)x \quad (5)$$

and adjusts his prices accordingly. Excess demand (supply) in a particular commodity means that the price must increase (decrease) before the next round of bidding. This continues until the *total* excess demand for all traders vanishes, which defines equilibrium. Trades are then allowed to occur. Most real markets do not behave this way. Adam Smith's main idea, the notion of a trader making a profit (financial friction), is completely eliminated from the price-adjustment process of GET. In the face of such criticism economists tend to reply: "You misunderstand the idea of equilibrium due to the balance of supply and demand. We only want to *prove* the *existence* of equilibrium", where, by "proof", they mean only *mathematically*, not empirically. In statistical physics, in contrast, the thermodynamic limit is generally impossible to prove mathematically but very good approximations to thermal equilibrium are accurately observed in careful, controlled experiments. An example is given by experiments that measure the specific heat at constant pressure to the fifth decimal place near a second order transition. This analolgy is important because the notions of utility and equilibrium were lifted historically from physics, partly from dynamics and partly from thermodynamics [3].

In the trial and error ('Tatonnement') price-adjustment process described above *stability* of equilibrium was assumed implicitly, otherwise no equilibrium is reached during the auction. In order to model Adam Smith's invisible hand, which must always move prices toward a stable equilibrium point in classical and neo-classical economics, we must invent a dynamical system and study stability. Invention, as opposed to nature with unique Newtonian dynamics, is not unique, so we are forced from the start to consider topological classes of dynamics by having to fall back on mathematical postulation rather than Galilean empiricism. The nontrivial difficulties facing anyone who might want to try to discover the correct topological dynamics class for a given market from statistics or time series analysis are discussed in [5].

## 2. Arrow-Block-Hurwicz dynamics

In order to discuss stability of the predicted equilibria the econometrician, with no Galileo-Kepler-Newton shoulders to stand on, must first invent a class of dynamics models.

In ABH dynamics [6] one starts as above by postulating a scalar utility function U whose gradient is assumed proportional to the price covector for n commodities

$$p_i = \lambda \frac{\partial U}{\partial x_i}. \qquad (5b)$$

The commodities are represented by a vector x in commodity space, where the entries defining x are the quantities of each commodity held, supplied, or demanded, depending on the context. Commodity space is implicitly taken to be flat, otherwise x is not a vector. One then has two separate spaces, price space and commodity space, and these are not yet thought of as a phase space in GET. Given price as a function of demand (or supply) x, one must then be able to invert the relationship (5) to find demand as a function of price, x = d(p). Given that one holds n commodities at time t, the usually assumed budget constraint [1,7] (Walras' 'law') says that the sums of prices times numbers of commodities held at time 0 must be the same as the prices times quantities of commodities held at any later time t. That is, the scalar product (in our inner product commodity space)

$$M = \tilde{p}x \qquad (6)$$

will automatically be conserved when we get around to building the dynamics. The excess demand $\varepsilon_i(p)$ is then formed in commodity space for one consumer, the $i^{th}$ consumer,

$$\varepsilon_i(p) = \tilde{p}x - \tilde{p}x_o, \qquad (7)$$

where $x_o$ represents the quantities of each commodity initially held before the 'Walras auction' begins. Having constructed the excess demand for n commodities for one consumer one then uses permutation symmetry to write down the total excess demand vector for each commodity for m consumers. This yields the total excess demand vector $\varepsilon(p)$. The ABH model simply assumes that the price changes obey

$$dp/dt = \varepsilon(p), \quad (8)$$

so that Walrus' auction has been replaced by a flow in p-space. Equilibrium is described as vanishing excess demand, and the process is still called Tatonnement.

That a flow should be called 'trial and error' appears peculiar at first sight, because what could be further from the notion of an auction than a flow? However, if we construct solutions to (8) using Picard's method of repeated approximations then we have a trial and error approach to the exact solution that is guaranteed to converge whenever a Lipshitz condition is satisfied by the excess demand. The solution, however, needn't be an equilibrium solution and generally isn't, even as the time goes to infinity (this is a not only a question of existence and stability of equilibria, but of basins of attraction).

The Walras condition (the budget constraint M = constant)) means that that the vectors p and $\varepsilon(p)$ are perpendicular. This constraint confines the motion to an n-sphere in price space. General Equilibrium Theory then tries to restrict the studies to excess demands that have a unique equilibrium, and mainly searches for those equilibria, which is rather boring mathematics with trivial, uninteresting (not to mention inapplicable) dynamics. The search for a globally-unique equilibrium was a way to try to justify the *universality* of Adam Smith's invisible hand, the notion that, given a free market, the same conditions should lead to equilibrium in Moscow, Mexico City, New York, London, and Frankfurt. In this restricted picture, trading only occurs *at* equilibrium. Aside from the fact that trading nearly never occurs at equilibrium in real markets, asking for universality of solutions of dynamics equations for arbitrary initial conditions is too high an expectation. Even if the ABH model were qualitatively applicable Moscow and New York could well belong to different basins of attraction (nonuniversality of effects of excess demand), and equilibria need not hold in either market.

With only two commodities integrability is guaranteed by elementary calculus. Integrable dynamics models with three commodities are discussed in the readable paper by Scarf [7]. Integrable models with three commodities have the nice property that the excess demand vector must be the cross product in price space of the gradients of two global functionally-independent conservation laws $G_1$ and $G_2$ [8],

$$\varepsilon_1(p) = \mu \nabla G_1 \times \nabla G_2 , \qquad (9)$$

where $\mu$ is Jacobi's multiplier. Scarf then shows that it is easy to construct model utilities where there is no approach to equilibrium by giving several nice examples, one of which has an unstable focus surrounded by a stable limit cycle [1].

An approach to equilibrium requires a driven dissipative system, which requires that the divergence of the excess demand vector in price space cannot vanish identically,

$$\nabla \bullet \varepsilon(p) \neq 0 . \qquad (10)$$

Smale's [9] 'global Newton method' of locating equilibria, if equilibria exist, implicitly uses the n-1 time-independent conservation laws $G_i$ of driven dissipative systems. The existence of n-1 local conservation laws for driven dissipative systems (integrable or not) was brought out in the open by Palmore [10], who was the first to emphasize that the damped simple harmonic oscillator has a global conservation law and actually constructed the invariant. Many of today's physicists and mathematicians seem to have trouble accepting the fact that even planar dissipative flows have a conservation law, although it was known to Jacobi and Lie and is trivial mathematics. In a driven dissipative system these conservation laws are singular at equilibria. Arnol'd [11] therefore refuses to label them as 'first integrals', but they are nevertheless conserved quantities, whether so-labeled by Arnol'd or not.

In a nonintegrable system the conservation laws exist only locally [5,8,11] so that the interactions can be transformed away only locally (the flow can be parallelized only locally), never globally. There are mathematical (but not economically realistic) conditions under which stable equilibria occur and are unique [1,6], but Scarf shows that it is easy to construct counter-examples where, e.g., an unstable equilibrium is enclosed by a stable limit cycle. Saari [12] points out that the class of more generally admissible excess demand vectors is so great that nothing, especially not deterministic chaos, is precluded. In other words, Adam Smith's invisible hand cannot be relied upon to match supply to demand globally. At best, it is a local phenomenon if it occurs at all in this theory. We now also know that flows in phase space may not be merely chaotic, they may be complex (no scaling laws, computer-like behavior, surprises at all length scales [5,13]). In other words, the invisible hand does not guarantee that equilbrium is stable, or even exists for an arbitrary excess demand in the allowed topological class, and there is no reason at all to believe in the universality of the effects excess demand on the basis of GET. As with all dynamics problems, the important effects are all local, not global. On the basis of econometric analysis, what works in New York need not work in Tokyo, Frankfurt, Paris, or Zurich. Or, what works for Microsoft need not work for Novartis. Some leading dynamical systems theorists have hoped that econometrics would be useful [14], and some have gone over to econometrics [1,15], but GET has never proven to be *empirically* useful.

Leaving aside the failure of the invisible hand to regulate the entire world in a trivial way, there are other more serious objections to the ABH model. First, what is the relaxation time for prices? This is a very good question because theoretical econometrics implicitly assumes extremely rapid relaxation to equilibrium. This seems strange to physicists, but we must remember that econometrics, especially GET, is more like a belief-system or logic-system than it is like physics or any other science. Unfortunately for physicists who might want try to do better, the left hand side of the ABH equation (8) has units of \$/sec, Euros/sec, or Yen/sec, whereas the right hand side (excess demand) is a pure number, and there is no way to discover a fundamental constant (a global invariant) with dimensions of money/time to repair this defect. Relaxation times simply cannot be predicted on the basis of the ABH model as it now stands. One could multiply $\varepsilon(p)$ by an interest rate to get the right units, but this would be arbitrary and meaningless.

Lying in the same boat is the question of the relaxation time of the commodity variables x. In the ABH model the prices are slaved to the commodity variables, which are implicitly presumed to relax infinitely fast: there are no differential equations

$$dx/dt = s(p,x) \quad (11)$$

for the commodity variables. A physicist might ignore the Walras law (the budget constraint), because physicists are usually pretty good at violating budgets anyway. Let's do this and look at a vaguely analogous mathematics problem in physics, a Hamiltonian system where x is position (phase space can always chosen to be flat [8]) and the covector p is the corresponding canonical momentum. If

$$p = \nabla U \quad (12)$$

where U is a scalar function of x, then the dynamics is globally integrable [8] and the motion is confined to an n-dimensional cylinder or torus in the 2n-dimensional (x,p) phase space. Einstein pointed out that assuming that p = grad U is equivalent to assuming integrability of the dynamical system [16], although this was shown earlier in Liouville's more complete explanation of integrability [17]. This is a formal non-economic analog of Walras' law. Here, there is no dissipation, and so there are no stable equilibria in phase space, only elliptic and hyperbolic points. For bounded motion one has eternal oscillation (with neutral stability). The function U is just the reduced action in this case. The condition for (12) (discussed explicitly by Liouville but not by Einstein) is that there are n global commuting constants of the motion, n global conservation laws of the Hamiltonian system. This then guarantees that Ldt is an exact differential in the x's and t, where L is the Lagrangian, so that the action is now a function (a generating function). The canonical momentum then is the x-gradient of the action or the reduced action.

If, on the other hand, the dynamics is nonintegrable, as is typically the case, then the reduced action

$$A_r = \int_C \tilde{p} dx \quad (13)$$

is a path-dependent functional, not a function, so that the analog of the utility U exists at best locally, not globally, and cannot even be written down as an infinite series without using analytic continuation. This might seem very far fetched as economics, but Hamiltonian systems do occur in econometrics (see section 3 below).

Hamiltonian dynamics shows that it is dangerous blindly to assume that the question of path dependence of a utility functional (13) can be divorced from the question of integrability of the underlying dynamics. If Mirowski [3] is correct about the history, then Gibbs tried but failed to make a similar point to I. Fisher. Gibbs presumably pointed out that the utility is generally path dependent, and suggested that Fisher address the problem of nonintegrability in relation to a possible underlying dynamics. Fisher did not understand nonintegrability of differential forms (neither did Walras or Pareto), so that Gibbs' point was apparently lost on Fisher, who erased all mention of nonintegrability from his last papers on utility after he had become established as an economist. We can speculate that Gibbs knew and understood Liouville's integrability theorem in Hamiltonian mechanics and had in mind precisely the discussion above. We don't know. We do know that Fisher's inability to address and resolve the question of the general path-dependence of utility left Samuelson and other economists to worry about nonintegrability some fifty years later. Most economists finally stopped worrying about it, but without having solved the problem. Others thought that merely writing down Slutsky relations ('Maxwell relations') was the answer. Mirowski made the mistake of trying to identify utility with potential energy, instead of action, in classical mechanics. The problem posed by Mirowski's stimulating criticism is solved next in part 3.

**3. Control via optimization in econometrics**

The Hamiltonian approach in econometrics allows us partly to repair two flaws in the ABH model in a simple way: a time scale can in part be introduced into the dynamics problem, and equations of motion for demand/supply ( or "production") variables x are introduced.

Hamiltonian systems fall out of optimization problems in control theory [18] because every variational principle formally yields a Hamiltonian system [19]. In econometrics the Hamiltonians that occur are not of the form studied by physicists, but many general theorems on Hamiltonian mechanics, especially Liouville's integrability theorem, apply independently of the form of the Hamiltonian. A Hamiltonian dynamical system is a group theoretic idea arising from a variational principle [20].

We begin with the discounted utility functional (the price of money is discounted at the rate $e^{-bt}$, e.g.)

$$A = \int e^{-bt} u(x,v,t) dt \qquad (14)$$

where $u(x,v,t)$ is the undiscounted 'utility rate' and v is a set of control variables or control functions ("instruments"). Optimize utility with respect to the 'set of instruments' v, but subject to the constraint

$$\dot{x} = s(x,v,t) \qquad (15)$$

where s is the production function (this is Mayer's problem in the calculus of variations [20]). The system (15) may be driven-dissipative in commodity space with the variables v held constant ((15) could be a damped linear oscillator or a Lorenz model, e.g.). Optimization yields

$$\delta A = \int dt (\delta(e^{-bt}(u + \tilde{p}'\delta(s(x.v.t) - \dot{x})) = 0 \qquad (16)$$

where the $p'_i$ are the Lagrange multipliers. The extremum conditions are

$$H(x,p',t) = \max_v (u(x,v,t) + p's(x,v,t)), \qquad (17)$$

$$\frac{\partial u}{\partial v_i} + p'_k \frac{\partial s_k}{\partial v_i} = 0, \qquad (17b)$$

(sum over repeated index k) which yields 'the positive feedback form'

$$v = f(x,p,t) . \qquad (18)$$

Substituting (18) into (17) yields.

$$H(x,p',t) = \max_v(u(x,v,t) + \tilde{p}\text{'}s(x,v,t))$$

$$\dot{p}' = bp' - \nabla_x H$$

$$\dot{x} = \nabla \cdot H = S(x\ p'\ t) \qquad (19)$$

where, with $v = f(x,p',t)$ determining the maximum in (17), $S(x,p',t) = s(x,f(x,p',t),t)$. The integral A in (14) is just the Action (the discounted utility rate is the Lagrangian).

It is easy to prove (19) by using by using the chain rule. To show that we actually have a Hamiltonian system, use the discounted utility rate $w(x,v,t) = e^{-bt} u(x,v,t)$ with $P = e^{\pm bt}p'$ to find

$$h(x,p,t) = \max_v(w(x,v,t) + \tilde{p}s(x,v,t))$$

$$\dot{p}_i = -\frac{\partial h}{\partial x_i}$$

$$\dot{x}_i = \frac{\partial h}{\partial p_i} = S_i(x.p.t) \qquad , \qquad (20)$$

which is a Hamiltonian system. Whether or not (15) with constant v's is driven-dissipative this system is phase-volume preserving, is conservative (and h is generally time dependent).

In (19) we have a time scale determined by the discount rate b,

$$\dot{p}' = bp' - \nabla_x H , \qquad (21)$$

although Hamiltonian term in the excess demand still has no clearly defined units, so this is not a complete fix. This dynamical system generally does not obey Walras' budget constraint. The Lagrange canonical momenta $p_i$ are called 'shadow prices' for the production or investment process. If we think momentarily of $p_i$ as 'money' and b as an interest rate, then $bp_i$ contributes to the total excess demand for money (negative interest $b < 0$ is possible, and was issued in special cases by the Bank of Japan during 1999). The entire right hand side of (21) must be thought of as excess demand $\varepsilon_i(p)$ for commodity $x_i$.

In this formulation the discounted utility rate $L = w(x,v,t) = e^{-bt} u(x,v,t)$ is the Lagrangian $L$,. Since the Hamiltonian h depends on time it isn't conserved, but integrability occurs if there are n global commuting conservation laws. These conservation laws typically do not commute with the Hamiltonian $h(x,p)$, and are generally time-dependent [8,17]. The integrabilty condition (due to n commuting global conservation laws) can be written as

$$p = \nabla U(x) \quad (12)$$

where for bounded motion the utility $U(x)$ is multivalued (turning points of the motion in phase space make U multivalued). U is just the reduced action given by (13), which is a path-independent functional when integrability (12) is satisfied (so is the action A given by (14)). When satisfied, the integrability condition eliminates chaotic motion (and complexity) from consideration because there is then a global, differentiable canonical transformation to a coordinate system where the motion is free particle motion (n commuting constant speed translations on a flat manifold imbedded in the 2n dimensional phase space). Conservation laws correspond, as usual, to continuous symmetries of the dynamical system (20).

Note that if integrability holds so that $U = A_r$ then p is not merely 'shadow price' but is, according to the economists, simply 'price'. There is no difference between calculating p as the gradient of A or $A_r$ in the integrable case. Either way, one gets the same price covector p. *However, the corresponding p is not an 'equilibrium' price covector because (12) is not an equilibrium condition,* in spite of the fact that the utility rate $w(x,v,t)$ is maximized with respect to the instruments v while maintaining the production rate constraint.

The equilibria that fall out of control problems in the 2n dimensional phase space of the Hamiltonian system (20) cannot be attracting (are not stable). By Liouville's theorem (which is all that's needed to prove Poincare's recurrence theorem whenever the motion is bounded) equilibria are either elliptic or hyperbolic points (sources and sinks in phase space are impossible in a Hamiltonian system). Bounded motion guarantees that there is eternal oscillation (stable or unstable), with no approach to equilibrium, because either the motion is neutrally stable with elliptic points as equilibria, or all equilibrium points are unstable (hyperbolic) and the motion may be chaotic or complex (no example of the latter has been constructed but Moore [13, 21] has suggested that it may occur for n = 3, whereas chaos requires only n = 2). For unbounded motion there must be at least one hyperbolic point. Integrable motion that is not oscillatory must be unbounded (see [22] for examples in econometrics). Econometricians still call hyperbolic points "stable" [22], by which they mean: choose initial conditions on the stable asymptote. To do this, Adam Smith's hand is presumed to be infinitely precise and certainly cannot be subject to any noise.

If we have an integrable system (20) and use (12) to find demand as a function of price, $x = d(p,t)$, then we can substitute back into the right hand side of the price equation in (20) to obtain

$$\dot{p} = \varepsilon(p,t) \qquad (8b)$$

which is not a volume-preserving flow in the in the n+1 dimensional price-time subspace of phase space because

$$\nabla_p \cdot \varepsilon(p,t) = \sum_{i=1}^{n} \frac{\partial \varepsilon_i}{\partial p_i} = -\sum_{i=1}^{n} \frac{\partial^2 h}{\partial x_i \partial p_i} \neq 0 \qquad (8c)$$

In other words we end up with a special case of non-phase volume preserving ABH dynamics. It's instructive for a physicist to solve the simple harmonic oscillator by this method: the equilibria that are defined by setting the right hand side of (8b) equal to zero are not oscillator equilibria in the full phase space, but only represent that the oscillator passes through the equilibrium position with either positive or negative finite momentum ((12) is not an equilibrium condition!).

Should central bankers and economic planners have faith in econometric models that are based on controlling a finite number of instruments $v_i$? No, because models with chaotic or even complex motions cannot be ruled out of consideration. All of the models have hyperbolic points as equilibria [22], which asks too much of Adam Smith's hand.

In formal welfare economics one assumes additive utilities [1]: one satisfaction function per "agent". An "agent" may be a consumer, the Fed, a factory, or a welfare recipient, depending on the context. Pareto efficiency is defined by maximizing the utility of one agent while holding the utilities of the other n-1 agents constant. A Pareto efficient allocation is purported to describe the best welfare distribution: each agent/recipient is supposed to be as well off as possible, given the utilities of other recipients. Pareto efficiency, now popular in legal circles [23], is only a mathematical model that generally does not describe reality. Utility functions don't exist in the absence of integrable dynamics, and integrability vs. nonintegrability cannot be decided in court. Some high courtroom decisions are, however, motivated by arguments from utility maximization.

The main point of this section could have been discovered by Sato [24], who applied Frobenius' theorem on integrability to equation (12) (this is somewhat analogous to cracking a nut with a sledgehammer). He did not apply similar mathematical horsepower to the Hamilton-Jacobi equation, which was discussed at the level of Goldstein or Landau-Lifshitz, rather than at the level of Whittaker, near the end of his book on the use of symmetries and Lie methods in econometrics.

4. **A toy model of the stock market**

Trading of commodities is postulated to occur in GET only at equilibrium where supply exactly matches demand, just like the pictures ('graphs') in Samuelson's text [2] (the 'graphs' in that text do not follow from real data or real theories, but are just sketches of expectations based on the unverified speculations of GET). At equilibrium agents hypothetically buy/sell reversibly, without any loss, in analogy with the reversible transformation of ice into water at zero degrees centigrade, or like the reversible transfer of random kinetic energy into work in a Carnot engine. Approximately reversible water-ice transformations are possible in carefully controlled experiments, but low financial-friction trading is usually possible only in very liquid markets over the shortest time intervals, at best. Markets in Europe are generally irreversible (sales are generally final). For example, Karstadt will not likely refund your money later in the afternoon after a morning purchase just because your wife doesn't like the color of the shirt you bought. Some markets in the US are reversible: if you buy a pair of Jeans at K Mart or Dillard's you can usually reverse the trade perfectly on a time scale of a few days (ignoring gasoline costs and wear and tear on the car). Most markets are not liquid enough to permit approximately reversible trading (low-friction trading), and whenever they are liquid the friction usually cannot be ignored. For example, You cannot immediately resell a new car to the dealer and get your money back after signing the contract. He will make a second profit from you, otherwise no trade will take place. The less liquid the market, the bigger the spread (cars, housing). Bond traders and specialists in stocks maintain bid/ask prices that represent small but non-negligible 'financial friction'. Discount brokerages offer low-friction trades in a liquid market. Liquidity seems to require finite and small but non-negligible friction. A trade likely will not be reversed if the seller would lose a big profit via the reversal.

Black [25] emphasized that noise is essential for liquidity. Noise and friction go together. Small friction (low transaction fees) is necessary for a liquid market. Finite friction limits noise at the highest frequencies (shortest time intervals) by placing a lower limit on price changes that can speculated over with any hope of profit. An enlightening toy model of the market, where prices change discretely, is discussed by Osborne [26] who constructs realistic examples of discrete supply and demand functions of price for traders placing limit orders in the stock market. Here, we find a nice example of econometrics as a good astronomer/physicist practiced it.

Following Osborne, we can (to zeroth order) understand the working of a stock market via the following model. Market orders are always executed, so we ignore them in favor of limit orders, which require a specialist (market maker) who tries to match supply to demand. If there were only market orders, then prices would often be determined by matching total demand to total supply, as in the economists' idea of 'equilibrium', but only in the absence of 'runs' on a stock. Limit orders prevent this 'equilibrium' from occuring even in orderly markets. Making trades possible at all is the stock specialist (market maker), who starts each day with a definite supply of a particular stock S and a definite supply of cash C. Both inventories are large enough to take care of 'normal' trading with a specified spread in bid/ask prices. How the specialist makes his profit via the spread is described by Osborne, as is the size of the inventory C. The job of the specialist is not to help to drive prices up or down, but to try to keep price changes small and to maintain as orderly a market as is possible, given excess supply and demand expressed by market and limit orders. Very liquid markets (heavily-traded stocks and bonds) have low financial friction whereas the transaction costs for housing (real estate broker) are high in a relatively viscous market.

Three prices should normally be shown on your discount broker's web site in real time: the last sale price, a bid price and an ask price. Assume for simplicity that trades are only made in round lots (100 shares). Suppose that a sell order for 200 shares of S is executed at the bid price of 20 1/8, which we can take to be 1/8 of a dollar lower than the ask price (typical bid/ask spreads may be 1/16 or 1/8 of a dollar on the NYSE, depending on liquidity of the stock S). The next bid and ask quotes will be lowered to 20 and 19 7/8. Correspondingly, if the last limit order executed was a buy of one or more round lots, then the bid/ask quotes are both increased by 1/8. If no one wants to buy at 20 and there is a mass of buy orders at 10, then either the bid/ask prices must be dropped dramatically or else trading in the stock must be halted for the day (like closure of a bank when there's a threat of a run). The stock market is not in equilibrium: the bid/ask and trading prices are not equilibrium prices. Even when one limit order can be filled (by matching a buyer at 20 1/8 to a seller at 20, e.g.), most limit orders cannot be filled: the total excess demand for the stock S does not vanish. Economists label stock market trading as 'equilibrium' because they tend to label everything as 'equilibrium', even when it is not. I discuss this in detail in part 8 below.

What about the 'real value' of an asset like a stock? Black [25] argues that 'value' is both random and unobservable, whereas price is random and observable. Believing strongly in his unstated, therefore undefined, notion of 'equilibrium', Black asserts that price always tends to return to value after large deviations, but we will see in the next section that the model used by Black and others to describe stock market movements has no restoring forces, only random walkers. Black also says that he will call the market 'efficient' if price is within a factor of two of 'value'. Since 'value is both unobservable and undefined, this assertion has no meaning. If we try to identify 'value' with price then we may fall into the trap of accepting the efficient market hypothesis (EMH), which states that dart-throwing in designing a portfolio is as effective trying to choose stocks on the basis of some other rational basis, like an estimate of 'value'. There is evidence that dart-throwing can be beaten over short enough time intervals that information has not had time to propagate to all investors (see [27,28], for example). Not all speculators/investors react the same at the same time to the same information, and investors (agents) typically are not rational but act on 'hunches', on the basis of someone else's advice, or (as Black points out) on the basis of noise perceived as information. Black observes that we never can be certain that we have information instead of noise. He used the now-popular term 'noise trader' and argued that 'information traders' can make money by timing their trades to take advantage of noise traders, who make up the bulk of the market and are responsible for the liquidity of the market. The EMH is a good assumption for agents who do not follow the news closely and who react too slowly to anticipate changes in market conditions. The EMH advises these agents to invest in index funds, and advises them not to follow the advice of 'financial advisors', who generally do not know any more about predicting market advances or declines in a stock or the entire market than does a dart thrower. In the end, the value of an asset like a stock or bond that you can't eat, sleep in or under, warm or cool yourself with, ride on, or use in any other desired way without liquidating it is worth, at any given time t, exactly what the highest bid in the market says it's worth at that time. It's price at any later time is unknown and undetermined. This does not prevent good traders from making money. Agents can forget about the illusions of 'value' and 'equilibrium' and concentrate on prices in the face of excess demand. This leads, quite naturally, to globalization, to the capitalization of everything everywhere. Whether or not it is a good idea is not discussed here.

For neo-classical discussions of liquidity traps, where demand can't be matched to supply although agents have money to spend and want to make trades, see Ackerlof [29] and Krugman [30].

All other things being equal, consumers prefer falling prices, or at worst equilibrium, in a market whenever they have to buy items for consumption or real use. People who buy stocks or offer credit prefer rising prices of assets. No one who invests in the stock market wants equilibrium (vanishing excess demand for the asset), because in that case there are no gains (due to dividends or price appreciation). A booming stock market is very far from equilibrium: the average excess demand is positive and large, otherwise prices can't increase. The US stock market from April-October, 1999, is roughly approximated by equilibrium (index funds like vfinx (Vanguard 500) have approximately no net gain/loss). During that time many individual stocks have taken big falls (cpq, one, tyc, ...) while others have skyrocketed (rhat, ge, ...). The fluctuations ('volatility') have been large for many individual stocks. Most individual stocks are not in equilibrium even when the market approximately is. Long-time expectations of typical investors for the stock market are of nonequilibrium (big gains), not equilibrium. No one would invest for the long haul in a stock market that is expected to be in equilibrium. Prices cannot rise unless the bidders dominate the askers, which is very good for the askers: they can unload risky assets that otherwise might produce no profit.

The stock market is generally not in equilibrium. Supply does not match demand because all limit orders cannot be filled. There is no 'clearing price' for one day, only a range of bid/ask prices over which trades were made with nonzero total excess demand during the entire day. The notion of equilibrium is not very useful and is even misleading in modeling financial data dynamically. I will oppose the tradition of the economists to call the stock market an example of equilibrium. Following the example set by Black [25], if they do not accept my argument and revise their description then I will attribute it to noise. Likewise for the misconception that utility maximization describes 'equilibrium'.

For an econometrician's approach to stock market pricing see O'Hara [31], where Osborne is not mentioned but utility is. For other recent attempts to model markets, particularly limit orders, see [32,34]. For a more recent discussion of noise traders see [34].

## 5. Noise and stochastic models of price changes

Deterministic models have not proven useful for describing stock, bond, and money market price motions. Evidence for deterministic chaos was not found in market data [35]. One can relax budget constraints to allow creation and annihilation of money, which is realistic. Money is not conserved. It is created and destroyed via credit, bond sales and recalls, defaults, etc. Money is created and destroyed with the tap of a computer key. A model that a statistical physicist may want to try to improve on is called the capital asset pricing model (CAPM). CAPM is based upon the 'law of one price' [36], meaning that arbitrage possibilities that occur over short enough time intervals have been acted on so that the economists' notion of 'price equilibrium' (equal assets have equal values) holds. The model uses random noise to represent the fact that we cannot know what the future price of each of n assets $S_i$ will, even after short time intervals. Economics and finance textbooks [2,36] teach 'price equilibrium', but trading houses live from arbitrage [37, 38].

Let p be the price of asset S at time t and let r be the rate at which the asset changes systematically in price, if there is a systematic change in price. Money-asset markets are modeled by an excess demand of the form

$$\frac{dp}{dt} = \varepsilon(p) = r(t)p + \eta(t) \qquad (22)$$

where r(t) is the drift rate for p and η includes everything else that contributes to the excess demand. The right hand side of (22) is the total excess demand for the asset. The net excess demand for S, $\varepsilon(p) = rp + \eta$, must be expected to be positive in order to attract investors since most people do not willingly throw away money. If r is constant and $\eta = 0$, then we have exponential growth of wealth of the asset compounded at interest rate r (nonequilibrium), as in a savings account or treasury bill over a period when interest rates don't change. Interest and credit represent excess demand for money (in Japan, in 1999, there is excess supply because people tend to save rather than consume). When r is deterministic and there is no noise then we have an example of a so-called 'risk-free asset' like a T-bill (this is a definition, not a description). If we try to apply this picture to risky assets like bonds or stocks (historically mislabeled 'securities') then r is not constant and is not determined in advance, and we may treat it as noisy so that (22) is a Langevin equation where both r and η represent noise. We can't calculate the future price p(t) of the asset because we don't know what r(t) and η(t) will look like over the investment period [0,Δt].

The earliest model of stock prices by Bachelier [39] sets r = 0 and takes η to be Gaussian random noise (this is a model of an equilibrium market because there is no expected price change Δp). A Gaussian distribution of η does not fit stock price changes. Osborne [39] created a revolution in finance by pointing out that one should instead study the variable x(t,Δt) = log(p(t + Δt)/p(t)) = log(1 +Δp/p), where x is Gaussian random noise with mean square fluctuation $\sigma^2 \Delta t$. Osborne found that stock prices could be fit approximately by this assumption with σ (for a given stock) constant, which would mean that stock prices are lognormally distributed. (Mandelbrot later found that cotton prices have a Levy distribution at large price increments and argued that σ is formally infinite). Osborne [39] argued that one should study logp on the basis of Fechner's law. Gunaratne [40] argued that we need an additive variable, and x(t, Δt) is additive, in order to apply the central limit theorem. I note also that one needs an additive variable on the left hand side of a Langevin equation because we take random noise to be additive (stochastic integration). The finance theorists then study the Langevin equation

$$dp(t) = p(t)r(t)dt + p(t)\eta(t)dt \quad (23)$$

where η is Gaussian random noise. From here on I will use the preferred language of finance and will use Doob's form [41] of the Langevin equation

$$\Delta p_i = p_i r_i(t)\Delta t + p_i \sigma \Delta B(t) \quad (24)$$

instead of the equation (23), because 'stochastic calculus' [41] based on the Ito lemma [42,43,44] allows one to derive Smoluchowski equations not only for p but also for functions w(p,t) of p, which leads to option pricing: w(p,t) may the price of an option to buy the asset S. Δp/p is the fractional change in price over a small but finite time interval $\Delta t$. Also, we can take $w = \log p$ to get the Langevin equation for $\log p$

$$\Delta \log(p_i(t + \Delta t)/p_i(t)) = p_i(t)(r_i(t) - \sigma^2/2)\Delta t + p_i(t)\,\sigma \Delta B(t) \quad (24b)$$

In (24) $r_i$ is Gaussian random noise with expectation $R_i$ and variance $\sigma_i^2$, while $\Delta B$ is Gaussian random noise with mean equal to zero and mean square fluctuation given by

$$\langle \Delta B(\Delta t)^2 \rangle = \Delta t \quad . \qquad (24c)$$

In other words, both p and x are assumed to do a random walk, including with respect to expectations about gain about some average expected return R (we can also replace (24c) by a Levy flight assumption to discuss fractional Brownian motion).

The beginning of the investment period for the portfolio of n assets $(S_1,\ldots S_n)$ is t, and the time horizon is $\Delta t$. The CAPM model calculates the total expected return R for the short investment period $\Delta t$, the sum of the fractional the price changes

$$R_i = \left\langle \frac{\Delta p_i}{p_i} \right\rangle = \langle r_i(t) \rangle \Delta t \quad , \qquad (25)$$

where the average is taken with probability density $P(r, \Delta B) = P(r) P(\Delta B)$, and where and $P(r)$ and $P(\Delta B)$ both describe uncorrelated Gaussian random noise. Economists refer to 'forces' that cause prices to return to 'value' after large fluctuations but 'value' is both undefined and unobservable, and there are no 'restoring forces' in this drunken sailor model (24) of stock prices. If the random walker happens to hit or come near any particular price a few times, at random time intervals, then that is just an accident that will likely happen 'if we wait long enough' for a statistically meaningful expectation value to be realized. However, no force, and certainly not Adam Smith's hand, has acted during the random walk, where we must think of the entire right hand side of (24) as excess demand. Excess demand is treated as random noise. If $R_i$ is the return on asset $S_i$, then the return at time t on the portfolio of n + 1 assets will be

$$r(t) = \sum_{i=0}^{n} x_i r_i(t) \qquad (26)$$

where $S_0$ is a risk-free asset ($r_0 = R_0$ is deterministic) and the other n assets are risky (the $R_i(t)$ are Gaussian random variables for i = 1,2,...,n). Next, one writes

$$x_0 = 1 - \sum_{i=1}^{n} x_i \qquad (27)$$

to obtain the (randomly fluctuating) portfolio return as

$$r(t) = R_0 + \sum_{i=0}^{n} x_i(r_i(t) - R_0). \qquad (28)$$

If we form the mean square fluctuation of (r(t) - $R_0$) and minimize it subject to the budget constraint

$$1 = \sum_{i=0}^{n} x_i \qquad (29)$$

and the constraint that the expected return at time t

$$R(t) = \sum_{i=0}^{n} x_i R_i(t) \qquad (30)$$

is fixed, then we arrive at the prediction (31) of the CAPM also and the definition of β. The portfolio is called 'mean-variance-efficient'. Given the expected return (31), minimizing the mean square fluctuation in the portfolio return is seen as minimizing the risk because the variance is here identified as the 'risk' [36]. Utility maximization is forgotten in favor of risk minimization.

Suppose there is an index fund that is known to be more or less efficient in this sense. Each of the n assets in this portfolio is risky. The expected return of asset $S_a$ in the portfolio is

$$R_a = R_o + \beta_a(R_e - R_o) \qquad (31)$$

where $\beta_a = \sigma_{ae}/\sigma_{ee}$, $\sigma_{ee}$ is the variance of the efficient portfolio and $\sigma_{ae}$ is the covariance of the risky asset with the efficient portfolio (which we may take to be the market itself, or an index fund). To try to understand the Fed Chairman's speeches, which can have a big effect on liquid asset markets during a speculative bubble, it helps to know what the phrase 'risk premium' means. The 'risk premium' is simply the second term on the right hand side of (31) [36].

It has long been known that the prediction (31) of CAPM does not agree with the data. The NYSE data show better returns for low-beta stocks than for high ones from the investment period 4/57-12/65 [45,46], which is the reverse of the model's prediction (higher returns for higher risk). This model is based on EMH, namely, that all assets do a random walk so that dart-throwing is predicted to be as good as any other method when it comes to choosing a portfolio of stocks. Another way to say it is that the Brownian motion approximation ignores short time intervals over which new information can cause price changes and during which arbitrage is possible. As Black points out, however, we can never be sure that the perceived information is not really just noise.

The biggest fault with the model is that the variances, as Mandelbrot showed (see also Malkiel [46] for later references), are not well-defined but show sudden sharp changes when computed over increasingly longer time horizons. In applying the model one can not use a long time series to obtain a 'global', estimate for the variance. One should instead estimate the variance locally for the period $\Delta t$ over which the investment is made. The variance estimates must be continually revised as time goes on. In the worst case one must expect 'surprises' (like the collapse of prices of Compaq, Tyco, Raytheon, Bank One, or any number of other observed large price drops during spring-fall, 1999) that are not be included in the CAPM estimate of risk (large deviations are ignored in the lognormal model). These 'surprises' are the real nature of complexity, and by their very nature [13] cannot be predicted by any model of the market.

The stochastic equation (24) is far from a complete dynamical model: it fails to predict R. For a thought-provoking discussion of the complexity of trying to understand how to try to model a prediction of R somewhat realistically, see [47]. No model explains why market expectations R are sensitive to small interest rate changes. Malkiel [46] produces a back of the envelope calculation that shows that this could be understood if dividends (instead of price increases) were the main thing causing agents to buy assets, but in a speculative bubble (like the US stock market from 1994-1999, e.g.) dividend expectations are likely negligible in comparison with expectations of capital appreciation due to rising prices.

In the CAPM we have obtained an incomplete prediction without having used utility. According to Varian [1] the CAPM can only be made consistent with the economtrician's notion of expected utility in two restricted cases. Given a probability measure/distribution µ(x), then if utility were a function we could define the expected utility as

$$\langle U \rangle_t = \int U(x,t) d\mu(x,t) \quad . \quad (32)$$

An example of a probability measure is the empirical measure defined by the data. Varian shows that optimization of expected utility will not reduce to the CAPM unless (I) the utility is quadratic, or else (II) all n assets are themselves normally-distributed. However, Merton has derived the entire picture from utility theory [48]. The main point is that we do not need utility to arrive at the CAPM, nor do we need utility to derive the Black-Scholes model. The expected utility functional is used in stochastic control theory, and gives rise to the diffusive 'Hamilton-Jacobi-Bellman' equation of stochastic dynamic programming theory [42,48,49]. However, the word 'utility' cannot be found in many other interesting and useful books on finance [36,43,44,50,51].

## 6. Black-Scholes Option pricing

Black and Scholes (B-S) give two easy derivations of their option pricing equation (for those who are bored by easy to follow derivations, see Merton [48]). The second derivation is the most enlightening, and provided Black's motivation [52]: the B-S equation is a straightforward application of the CAPM. This viewpoint is most useful, as it shows that modification or falsification of one model requires the immediate modification of the other.

An option to buy an asset is the right to buy the asset at a specified price P within a specified time horizon T (the option expires at time T if not exercised). A 'European' call option can only be exercised at time T and will be exercised only if p - P ≥ 0, where all financial friction (brokerage costs) are ignored in this discussion. For $0 \leq t < T$ the option price w(p,t) obeys $P \leq w \leq p$. The question is: for $0 \leq t < T$ how should the option be priced? The future condition at t = T holds,

$$w(p,T) = \{p - P \text{ if } p \geq P, 0 \text{ if } p < P. \quad (33)$$

The idea is to apply the CAPM to two assets, a stock S and it's option $O_s$. Following Black and Scholes [53], note first that since for small price changes

$$\Delta w = \frac{\partial w}{\partial x} \Delta x . \quad (34)$$

This yields

$$\frac{\Delta w}{w} = \frac{\Delta p}{p} \left( \frac{p}{w} \frac{\partial w}{\partial x} \right) \quad (35)$$

so that

$$\beta_w = \frac{p}{w} \frac{\partial w}{\partial x} \beta_p . \quad (36)$$

Let

$$\Delta R = R_e - r_o \quad (37)$$

denote the difference between the expected return on the efficient portfolio (which we can take to be the market itself, or an index fund) and the interest rate on risk-free assets. β∆R is the risk premium.

In the CAPM we have the expected returns for the stock and option are given by

$$\left\langle \frac{\Delta p}{p} \right\rangle = R_p \Delta t = r_o \Delta t + \Delta R \beta_p \Delta t$$

$$\left\langle \frac{\Delta w}{w} \right\rangle = R_w \Delta t = r_o \Delta t + \Delta R \beta_w \Delta t. \quad (38)$$

According to Ito calculus [43], and because

$$\Delta p^2 = \sigma_p^2 \Delta t, \quad (39)$$

for short space-time intervals we have

$$\Delta w = w(p + \Delta p, t + \Delta t) - w(p,t) = \frac{\partial w}{\partial p}\Delta p + \frac{\partial w}{\partial t}\Delta t + p^2 \sigma_p \frac{\partial^2 w}{\partial p^2}\Delta t. \quad (40)$$

Taking the expectation value of this equation divided by w and combining with (38) yields the B-S equation

$$\frac{\partial w}{\partial t} = r_o w - r_o p \frac{\partial w}{\partial p} - \frac{1}{2} p^2 \sigma_p \frac{\partial^2 w}{\partial p^2}. \quad (41)$$

This is a backward-time diffusion equation. The final price w(P,T) and expiration date T are fixed, so the equation must be solved by backward integration in t to find the solution (forward time solutions do not exist). This derivation is not exhibited in popular texts [37,38] but is very enlightening because it shows that the limitations on the CAPM and B-S options pricing are the same. Both can, at best, be used over short enough time intervals $\Delta t$ that there are no drastic changes (surprises) in expected returns and variances due to shifts in agents' expectations in the market. The models can't be extended to long times without empirically revising the estimates for gain and variance as surprises occur. See [44] for the use of 'implied volatility'.

Like the CAPM, the predictions of the B-S model are wrong: out of the money options are systematically priced too low by the model,

and in the money options are priced too high. The fault lies with the lognormal assumption: the lognormal distribution is too fat near the peak and too small in the wings to fit stock prices accurately. Traders claim [54] that the model does not work at all for stocks (individual stock price distributions are themselves too noisy) but works for bonds and foreign exchange if 'smile' is introduced. 'Smile'(or 'frown') is the assumption that σ varies with p, which violates the model. 'Smile' is the engineers' way of fudging the parameters to make the predictions of the model 'work'. An improvement was discovered by Gunaratne [40], who found that the exponential distribution

$$p(x,t) = \begin{cases} Ae^{\gamma(x-\delta)}, & x < \delta \\ Ae^{-\nu(x-\delta)} & x \geq \delta \end{cases}$$

(41b)

holds for *all* price ranges in x(t,Δt) of the log of price increments for bonds and foreign exchange (individual stock distributions were too noisy to draw conclusions), where $\delta \propto t$ [40]. Littel-known is that an option pricing model for the exponential distribution is also derived in [40], which was never published because the work was done in a trading house rather than at a university. For other attempts to go beyond Black-Scholes see nearly any recent issue of the International Journal of Theoretical and Applied Finance.

A Hurst exponent, consistent with Δx as a Levy flight, has been extracted from market index [55,56] and foreign exchange [57] data. The exponential distribution can be distinguished from a Levy distribution if the price information is accurate enough that terms O(logx) can be detected in data analysis.

Option pricing with absorbing boundaries was solved for the exponential distribution by Gunaratne [58].

**7. Deterministic chaos vs. random noise at short times**

The main point built into a Wiener process is that at the shortest times stock prices are completely unpredictable: the price is everywhere continuous but has no derivative. This is consistent with short time unpredictability of real prices, which change in discrete units over very short discrete time intervals.

A flow in phase space has a completely opposite short time behavior. Chaotic and complex dynamical systems found in nature are not predictable at long times (mathematical models are still represented by computable dynamics so long as all parameters used in the model are computable), but at short times the flow is trivial: any flow can always be parallelized ('rectified') over a short enough finite time interval. Over a short enough time interval there is always a coordinate system (reachable via a differentiable transformation) where the motion is free particle motion [11] (local integrability). Chaotic systems are pseudo-random over long time intervals but are models of perfect predictability over the shortest time intervals. One can never 'derive' (23) from a phase flow like (8), although it is possible to use a system that is chaotic to generate discrete pseudo-random time series at long times. However, short-time integrability of phase flows like (8) and (19) is violated in the worst possible way by stochastic dynamical models. I have argued elsewhere that driven-dissipative chaotic deterministic dynamics generally cannot be replaced by stochastic dynamics [5].

Another way to emphasize the incompatibility of the deterministic and random models is to note that diffusion equations generate no characteristic curves [18]. The characteristic curves are the particle picture (for the same reason there is no particle, or wave, description of the Schrödinger equation in quantum mechanics). First order partial differential equations, linear and nonlinear, are equivalent to particle systems but the only second order differential equation that generates characteristic curves is the wave equation [18].

Generally speaking, in finance theory one could forget the entire elaborate framework of GET except that books by theorists like Merton still attempt to 'save the appearance' by trying to use the elaborate language of GET to describe far from equilibrium phenomena ('epicycles' by a new name). Black, on the other hand, advised experimentation and did not expect econometrics to survive in it's neo-classical form [59].

## 8. Nonequilibrium and instability in stochastic dynamics

According to economists predictions based on (24) are called 'equilibrium' even when the average excess demand does not vanish (R = 0 is not guaranteed and is even undesirable from an investor's standpoint). Even worse, they speak of 'market forces' that return the prices 'to value' (meaning, presumably, 'equilibrium') in spite of the fact that the drunken sailor described by (24) feels no restoring force whatsoever and may even wander far from equilibrium. A restoring force could only be introduced by letting the average drift R depend on price, by making a price potential well, e.g. Financial markets are instead modeled by a collection of n drunken sailors (assets) with constant average drifts R. The job of the Fed Chairman is to try to use words to restrain any collective drift of the drunken sailors, and also via interest rates and the money supply.

Let me describe the 'equilibrium' assumed to exist by the economist. The usual argument says to subtract out the systematic return R and study the fluctuations about that 'equilibrium', which is supposed to be described by a rescaled price

$$\tilde{p} = pe^{-Rt} \qquad (41)$$

that should satisfy

$$\frac{d\tilde{p}}{dt} = 0 . \qquad (41b)$$

in some sense. Such rescaling cannot be performed for an individual stock because for stocks R is unknown and is not even well-defined. The standard argument is that we can do this rescaling for bonds and foreign exchange, for the market as a whole, or for an index fund. I will now show that rescaling does not lead to the idea of equilibrium even for a low-risk asset, one where r is deterministic. Take the simplest case of a constant interest rate r over some time horizon t, for a treasury bond, e.g.,

$$\frac{\Delta p}{p} = r\Delta t + \sigma \Delta B(t) . \qquad (42)$$

Stochastic integration then yields

$$p(t) = p_o e^{(r - \sigma^2/2)t} e^{\sigma \Delta B(t)}, \quad (43)$$

so that we can rescale the price by the deterministic/systematic part to obtain

$$\tilde{p}(t) = p(t) e^{-(r - \sigma^2/2)t} = p_o e^{\sigma \Delta B(t)}. \quad (43b)$$

We can study stability by calculating the moments of the Gaussian distribution $P(\Delta B)$, and obtain

$$\langle \tilde{p}(t)^k \rangle = = p^k_o \langle e^{k \sigma \Delta B(t)} \rangle = p^k_o e^{k^2 \sigma^2 t/2}. \quad (44)$$

Clearly, this is not an 'equilibrium' distribution because, even if one chose a different scaling factor, e.g., in order to make the second moment stable, then the rescaled higher moments will all still diverge as t increases. The idea of defining statistical equilibrium by rescaling prices is not possible. Another way to see that we are not in equilibrium is to calculate the entropy of either the lognormal, or observed exponential distribution (41b) (where $\delta \propto t$), and observe that the entropy of the distribution

$$S(t) = - \int p(x,t) \log p(x,t) \, dx \quad (45)$$

increases with time. An equilibrium distribution would maximize the entropy subject to some constraint, which would allow the analog of the inverse temperature to be introduced to describe hypothetical 'reversible trading' due to the prevalence of lots of noise with low financial friction. In the absence of price potential wells, the drunken sailors can maximize the entropy and achieve equilibrium only if we introduce walls (upper and lower bounds on prices, or government controls) to confine the sailors. Negative feedback might stabilize the random walkers, as the Fed Chairman has hoped, but there is no evidence in the data that this psychologic approach can always be relied on.

The origin of the economists' misconception of stochastic dynamics as market 'equilibrium dynamics' follows from the misconception that maximizing the expected utility yields equilibrium. It doesn't, whereas maximizing the entropy would.

The idea that the bond and foreign exchange markets obey the exponential distribution seems at odds with the idea that markets are complex. In a complex system the distribution cannot be known in advance [13]. One can only watch to see how events unfold and then record them, but this unfolding gives no information about future surprises. Complexity could only arise in a market obeying exponential statistics via 'surprises', like changes in agents' expectations, that would violate the exponential distribution. That is, if agents should become aware that they are creating an exponential distribution and then try to exploit that information, our expectation is that the distribution of price changes x will change.

To the extent that prices are exponential, not lognormal, standard stochastic optimization theory will not work. A second order Hamilton-Jacobi-Bellman diffusion equation cannot be derived to describe the dynamics.

Notice also that for finite and large Δt we have on the basis of (42) that

$$\left\langle \frac{p(t + \Delta t) - p(t)}{p(t)} \right\rangle = e^{R\Delta t} - 1 \quad (45)$$

so that the CAPM prediction would hold only for RdΔt << 1 even if the lognormal model were correct. An improved CAPM based on the exponential distribution will be presented later.

A positive expected gain R can be a destabilizing factor that may actually produce increasing prices if agents act on the same expectations. According to the empirical data there are no internal 'restoring forces' that can stabilize a free market (the Fed Chairman implicitly acts on this assumption by trying to 'talk down' the speculative bubble of 1995-'99)). *Adam Smith's hand simply does not exist in this picture, where the balancing of supply with demand occurs randomly, infrequently, and only by accident.* As Lewis points out [37], a brokerage house is a 'full-service casino'. Unlike Las Vegas and Monte Carlo, they will not only accept your bets but will even lend you the money to bet with (margin trading)! Margin is a dangerous form of credit, so beware the gamblers' ruin (for an example of the gamblers' ruin in action, see the history of LTCM [38,60]).

Merton stresses, in writing down (24), that he assumes that all agents/investors have the same expectations about the dynamics of the market. Bubbles and crashes are caused by uniform expectations of many agents (with adequate capital) acting relatively coherently. Restrictions that make capital more expensive for speculation discourage expectations, causing bubbles to deflate. This is the hope of the Federal Reserve Chairman late in 1999. This is all in keeping with the idea that social behavior is not like natural law [5], but is actually created to some extent by expectations. Krugman [30] sees runs on a currency as an example of self-fulfilling expectations.

Black, who wrote beautifully and clearly like a very good physicist, was a firm believer in 'equilibrium', but his idea of 'equilibrium' apparently was not that of econometrics. He regarded the B-S equation as an 'equilibrium' model, so that the market is 'in equilibrium', for Black, when the model prices the options correctly (markets are not lognormal, but we can take the corrected B-S formula of [40] as the standard here). Black's idea of 'equilibrium' threfore seems more to corespond to the EMH, to the idea that arbitrage possiblilites are not present because information traders have eliminated them (so that 'the law of one price' holds temporarily), than to any dynamic idea of equilibrium that we have discussed in parts 1- 3 above. That is, I think that where Black writes 'equilibrium' we must substitute efficient market/no arbitrage possible.

## 9. Econo-logic vs empirical data

"Poets are the unacknowledged legislators of the world....Let those who will, write the nation's laws, if I can write it's textbooks."
(P. Samuelson, quoted by Berstein [61])

Is econometrics scientific? Since Galileo, the abstraction that constitutes a science must be empirically based. Forgetting for the moment about dynamics, can the idea of a utility function with (2) be used to predict or even describe demand, supply, and prices in real markets, or in somewhat realistic toy models of markets when we look at empirical data? Osborne [26] asked whether there is evidence that supply-demand curves pass this basic test. The evidence for econometrics, if it exists, does not appear in Samuelson's text. The supply-demand 'graphs' in Samuelson are just 'cartoons' that are not based upon any known empirical data. Economists like to discuss elasticity (requiring a continuously differentiable utility) but do not show that utility can be measured and do not estimate it or it's slopes (marginal utilities) from real data. There is, as Osborne pointed out, a very good reason for this. Given a hypothesized abstract utility (purely theoretical, not empirical), one could then use (2) to derive the predicted price vector p for n commodities as a function of demand or supply x, $p = f(x,\lambda)$. *Osborne points out that this relation cannot be extracted from real data*, and observes that there is no unique price as a function of demand or supply, calling into question (2) as the basis for *anything* . Nonintegrability of a Hamiltonian system yields exactly such behavior [16]. Example: given twenty tomatoes (supply), all other things being equal, then what's the price? Answer: anything or nothing. Question, given demand for 50 Ford Mondeos/Contours, what's the price? Answer: not able to decide it empirically (nearly twice the price in Germany as in America, in fact, but arbitrage is not attractive due to taxes and shipping prices, i.e., due to 'financial friction'). Osborne's observation is bad news for econometrics: it implies that the idea of path-independent utility makes no sense empirically (writing down utility as a functional may make sense whenever there is something to optimize, if we know the dynamics in advance). Osborne illustrates that demand and supply as a function of price do seem to make sense empirically and gives as examples shopping for dresses and filling market limit orders. Black [25] later pointed out that discrete data on demand as a function of price are extremely noisy, so that in practice the required curves can't be constructed from known empirical data.

## 10. Conclusion

I have shown that utility generally does not exist as a function, that utility maximization is not an equilibrium condition, and therefore that price generally cannot be expected to be defined as a function of demand variables. This means that standard econometrics is wrong, inapplicable (no great surprise to anyone, presumably) and cannot form the basis for intelligently-made socio-economic policy decisions. I point out further that market models are not equilibrium models, that there is no stability either in the models nor in asset price data. The entropy can be defined, but it is never maximized in an unregulated free market. Equilibrium does not prevail, and is not a useful idea for understanding real markets. As we well-know since the collapse of LTCM, the market cannot be counted on always to be relatively 'efficient' either: surprises occur in the form of large, unanticipated deviations from all model predictions.

Econometricians sometimes admit that their field is more like logic and pure mathematics than it is like an empirical science like physics [14,62]. Econometrics has a history of contributions by pure mathematicians, with very little input by physicists (see [26] for an attempt by an astronomer). Theoretical finance, through Osborne, Mandelbrot, Black, Scholes, and Merton, although still too much tied to Osborne's (wrong) lognormal distribution (see any text on finance), has developed from very different initial conditions, has developed more as an attempt to be an empirical science. But can central bankers, corporations, trading houses and hedge funds be expected successfully to turn Moscow into Wall Street and Chihuahua into Wal Mart? Neither the ABH model nor stochastic control theory provide any theoretical foundation for that expectation. Globalization, the capitalization of everything everywhere, is a very large uncontrolled experiment where no one knows the outcome. Events like the collapse of the USSR, the financial crises in Mexico and the Far East, the financial crisis in Brazil, and the collapse of Long Time Capital Management (LTCM) are examples of 'surprises' that were not predicted by any theory and which were not anticipated by many very astute financial agents (however, see [38] for Scholes' warning to LTCM before it crashed). The fact that markets operate on the basis of noise and complexity rather than on the basis of Adam Smith's controlling/stabilizing hand means that anything can happen, including long runs of either pleasant or unpleasant events. For a history of the evolution of the idea of Adam Smith's hand, see [63].

Black [59] points out that econometric theorizing is not accepted on the basis of experiment, but because researchers persuade one another that the theory is 'correct and relevant'. This, alone, is not enough to establish a theory in physics where precise identical repeated, experiments are performed (or in astronomy where there are no controlled experiments but careful observation provides accurate data), although string theory and various too-ambitious models in cosmology are open to some criticism on this count. Black expected that experiments eventually will be done in economics and finance whenever the desire is great enough. LTCM was a completely uncontrolled experiment. LTCM suffered the gamblers' ruin because the usual econometric expectations about arbitrage and market efficiency proved wrong: certain bond interest spreads widened instead of returning to (Black's idea of) 'equilibrium', as 'the law of one price' assumes they should. The problem here is that the decay of small deviations from thermodynamic equilibrium forms the model for what the economists expect, but there is no correct thermal equilibrium analogy for real markets, which are complex, far from equilibrium stochastic dynamical systems. Complex deterministic dynamics has been investigated [13,21], but complex stochastic dynamics remains an unvisited frontier.

In qualitative agreement with Black, a former dynamical systems theorist who worked in econometrics [15] argues that econometrics is socially-constructed and notes that the notion of market equilibrium was philosophically soothing in a time when conservatives and others were afraid of revolt and revolution by the masses. General Equilibrium Theory is more like a mathematics-based ideology than like a science. In the age of complexity it will not likely survive whereas Newtonian mechanics not only survived but generated the field of deterministic chaos, and has been speculated to contain complexity as well [13]. The recent deregulation of banking/insurance/brokerage, another uncontrolled experiment in finance, combined with the lack of regulation of options trading will likely lead to surprises.

**Acknowledgment**

I'm grateful to Kevin Bassler, Cornelia Küffner, Larry Pinsky, and Johannes Skjeltorp for criticism and discussions, to Kevin Bassler, Gemunu Gunaratne, and George Reiter for reading and criticizing parts of the manuscript, to Yi-Cheng Zhang, Arne and Johannes Skjeltorp, and Gene Stanley for encouragement, and to the following colleagues for specific references: Kevin Bassler for Varian, George Reiter for Mirowski [3], which generated my motivation to write this article, Gemunu Gunaratne for [40] and [58], Hugh Miller for Saari [12], Johannes Skjeltorp for Karatzas O'Hara [31], and Yi-Cheng Zhang for [27], Ackerlof [29], and Lo [33]. A forerunner of parts 1- 3 of this article appeared as a Nov. 1999 Econophysics Forum Feature "Is Econometrics Science?" Thanks also to Mike Marder who suggested in 1996 that it would be interesting to analyze how many of the graphs in Samuelson's text "Economics" can be verified by real data. We did not know at the time that Osborne had already begun that anaylsis.